\documentclass[5p]{elsarticle}
\pdfoutput=1

\usepackage{graphicx}
\usepackage{dcolumn}
\usepackage{bm}
\usepackage{hyperref}
\usepackage{color}
\usepackage{amsmath}
\usepackage{amssymb}

\def\slash#1{#1 \hskip-0.65em /}
\newcommand{\newc}{\newcommand}
\newc{\etmiss}{\slash{E}_T}

\begin{document}

\begin{titlepage}

\title{Multi-hadron final states in RPV supersymmetric models with extra matters}

\author[UB]{Masaki Asano}
\author[UL]{Kazuki Sakurai}
\author[UT]{Tsutomu T. Yanagida}

\address[UB]{Physikalisches Institut and Bethe Center for Theoretical Physics, 
Universit\"{a}t Bonn, Nussallee 12, D-53115 Bonn, Germany}
\address[UL]{Department of Physics, King's College London, London WC2R 2LS, UK}
\address[UT]{Kavli Institute for the Physics and Mathematics of the Universe (WPI), TODIAS, University of Tokyo, Kashiwa 277-8568, Japan}

\begin{abstract}
The gluino mass has been constrained by various search channels at the LHC experiments and the recent analyses are even sensitive to the cases where
gluinos decay to quarks at the end of the decay chains through the baryonic RPV operator.  
We argue that introduction of extra matters, which is partly motivated by cancelling anomalies of discrete R symmetry,
may help to relax the gluino mass limit when the RPV hadronic gluino decays are considered.
In the scenarios where the extra matter states appear in the gluino decay chains, the number of decay products increases
and each jet becomes soft, making it difficult to distinguish the signal from backgrounds.
We investigate the sensitivity of existing analyses to such scenarios and demonstrate that the gluino mass limit can be relaxed 
if the mass spectrum reconciles the sensitivities of high $p_T$ jet searches and large jet multiplicity searches.
\\~\\
{\small  IPMU14-0115, KCL-PH-TH/2014- 24, LCTS/2014-22 }
\end{abstract}

\end{titlepage}

\maketitle

\section{Introduction}

The first phase of LHC operation have achieved  
the intensive and comprehensive new physics searches.
The searches have so far only seen agreement between the Standard Model (SM) and data, which in tern placed stringent constraints on beyond the Standard Models (BSMs).
Especially, R-parity conserved minimal supersymmetric SM (RPC MSSM) is severely constrained due to luck of events in the large missing energy channels,
and the gluino mass in RPC MSSM is constrained up to about 1~TeV~\cite{ATLAS:gluino, Chatrchyan:2014lfa}.

The stringent limit on the gluino mass in the MSSM can be modified if R-parity violation (RPV)
is introduced%
\footnote{
The gluino mass limit can also be relaxed in the models with compressed SUSY mass spectrum (see $e.g.$ \cite{Alwall:2008ve}).
}
\cite{Allanach:2012vj, Evans:2013jna}.
In this case the LSP can decay promptly into visible particles, trading the large missing energy signature with large multiplicity of jets and leptons.
The RPV scenario which is the most difficult to be searched for would be the one where the pair produced SUSY particles decay fully hadronically via the $UDD$ 
baryonic RPV operator. 
Some of the recent ATLAS and CMS analyses however explicitly target such models 
and if the gluinos decay into three or five quarks, the six and seven jet analyses \cite{CMS:TJResonances, ATLAS:RPVgluino}
exclude the gluinos lighter than 900~GeV.   

In this paper we point out that the RPV scenario with extra matters 
may lead to event topologies where the gluino mass limit is more relaxed.
The extra matter scenario is one of the interesting possibilities of the MSSM extensions.
An advantage is that anomaly of discrete R-symmetry, $Z_{NR}$ $(N>2)$, in the MSSM can be cancelled 
by the extra matter fields.
For instance, it is known \cite{Kurosawa:2001iq,Asano:2011zt} that introduction of a $\bf{5} + \bf{\overline{5}}$ or a $\bf{10} + \bf{\overline{10}}$ chiral multiplet pair, or three pairs of  $\bf{5} + \bf{\overline{5}}$ can achieve non-anomalous discrete R-symmetry.
The discrete R-symmetry may play an important role in low-energy supersymmetry (SUSY) models.
It controls dangerous proton decay operators as well as the constant term and supersymmetric $\mu$ term in the superpotential. 
The mass terms of extra matter states are also controlled by the discrete R-symmetry.
For example, the mass terms with the similar scale to the soft SUSY breaking scale can be
generated by the Giudice-Masiero mechanism~\cite{Kurosawa:2001iq, Asano:2011zt, Giudice:1988yz}.

If several extra matter states involve in the gluino decay chain, the number of final state particles becomes large and $p_T$ of each visible particle tends to be small 
because the initial gluino mass energy is divided into a large number of decay products.
The sensitivity of the current RPV SUSY searches then drops because of the high $p_T$ jet requirement.
The sensitivity of the existing analyses to the hadronically decaying gluinos with long cascade decay chains has been studied in Ref.~\cite{Evans:2013jna}.
The authors pointed out that the CMS black hole search~\cite{Chatrchyan:2013xva}
can effectively constrain the light gluinos in this scenario.
We will demonstrate that the limit from the CMS black hole search can also be relaxed if there is a mild mass degeneracy between the gluino and
the LSP. 

The rest of the paper is organised as follows:
the next section  describes our model setup that includes extra matters and some RPV operators. 
In section \ref{sec:gluino-chain}, we discuss the gluino decay chains in the RPV extra matter scenario.   
In section \ref{sec:constraints}, we reinterpret the existing analyses and study the gluino mass bound in the context of simplified models. 
Section \ref{sec:discussion} and \ref{sec:conclusion} are devoted to discussion and conclusion.

\section{Extra matter with R-parity breaking operator}

We consider models with extra $\bf{5} + \bf{\overline{5}}$ chiral multiplet pairs. 
We write ${\bf 5}^\prime_i = (D^\prime_i, L^\prime_i)$ and ${\bf \overline{5}}^\prime_i = (\overline{D^\prime}_i, \overline{L^\prime}_i)$
and 
introduce the mass terms for these component fields.%
\footnote{
We use a notation in which $D$, $U$ and $E$ represent the chiral multiplets 
containing anti-particles. 
}
\begin{eqnarray}
W &\supset& 
M^{L^\prime}_i L^\prime_i \overline{L^\prime}_i
+  M^{D^\prime}_i D^\prime_i \overline{D^\prime}_i   ,
   \label{eq:Extramatter_mass}
\end{eqnarray}
where $M^{L^\prime/D^\prime}_i > M^{L^\prime/D^\prime}_j$ for $i < j$.
In order to have multi-step hadronic gluino decays we introduce the RPV operators
\begin{eqnarray}
W &\supset&  
   \lambda^{''}_{212} U_{2} D_{1} D_{2}  +
            \tilde{\lambda}_{i}^{L^\prime} L^\prime_i Q_2 D_2 ,
              \label{eq:Extramatter_RPV}
\end{eqnarray}
The first term is necessary to have the LSP decay into three light flavour quarks,
whilst the second term is needed to make the gluinos decay to $L^\prime_i$
as well as to make $L^\prime_i$ decay to a lighter $L^\prime_j$ ($i < j$) or a neutralino lighter than $L^\prime_i$ as we will see in the next section. 
Here and throughout the paper, $L^\prime_i$ and $D^\prime_i$ represent the superfields and/or fermionic components of the chiral multiplets
and assume that the scalar components acquire the soft masses and are heavy enough not to contribute to our analysis.

We assume $D^\prime$ is heavier than the gluino for simplicity and do not consider $D^\prime$ production.
If $D^\prime$ is light enough to be copiously produced at the 8~TeV LHC or appear in gluino decay chains,
the RPV operators such as $D^\prime_i D_j U_k$ would be required.

Although the baryon number is violated by the $UDD$ operator in our model, the lepton number is still conserved 
by declaring $L'$ fields have zero lepton number.
The proton decay constraint is thus avoided.
To satisfy the constraint from $n$-$\overline n$ oscillation and suppress the single squark production
we assume $\lambda^{''}_{212} \sim  10^{-3}$.

\section{Gluino decay chain}\label{sec:gluino-chain}

In the models where the gluino is the LSP and the $UDD$ operator is introduced,
gluinos decay into three quarks: $\tilde g \to 3 q$.
If there is a neutralino below the gluino, the gluino can decay into five quarks through the neutralino: $\tilde g \to qq \tilde \chi_1^0 \to 5 q$. 
For both cases, gluinos are severely constrained by the six and seven jet analyses \cite{CMS:TJResonances, ATLAS:RPVgluino}
and the limit on the gluino mass is found at 900~GeV.  

\subsection{Gluino $\to$ seven-quarks} 
\begin{figure}[t]
  \includegraphics[width=\columnwidth]{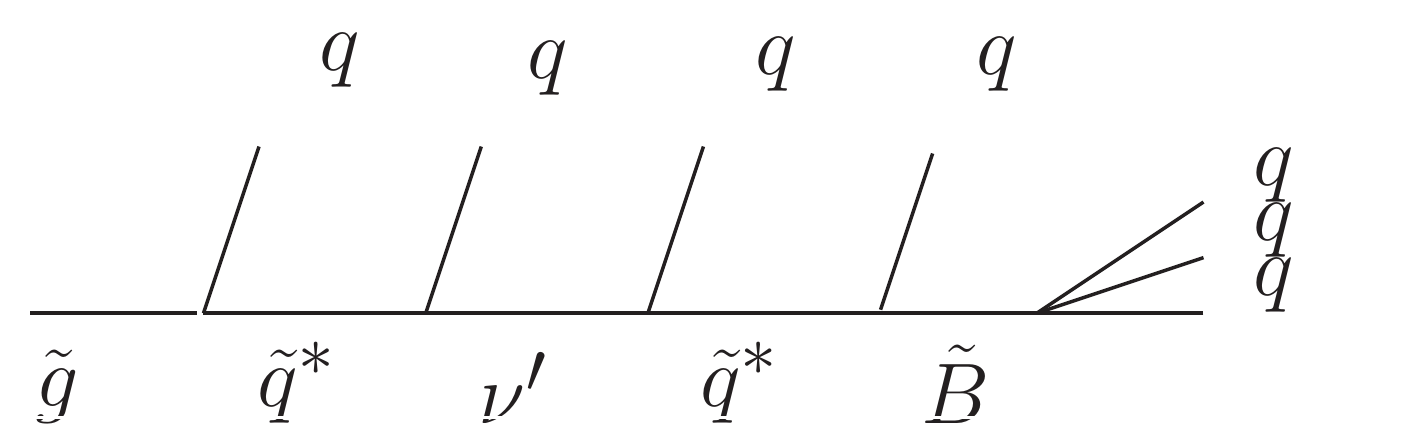}
  \caption{Typical gluino decay chains which are induced by adding one pair of $\bf{5} + \bf{\overline{5}}$ extra matter multiplets with additional RPV terms.}
  \label{fig:decay_diag1} 
\end{figure}

If the fermionic component of $L^\prime = (\nu^\prime, l^\prime)^T$ is lighter than the gluino,
the gluinos can decay into two quarks and $L^\prime$ via off-shell squarks through the $L_i^{'} Q_j D_k$ operator with the $\tilde{\lambda}_{ijk}^{L^\prime}$ coupling.
The $L^\prime$ can then decay into two quarks and a neutralino by the same mechanism as the gluino decay, $\tilde g \to qq L'$, if the neutralino is lighter than $L^\prime$.
The neutralinos finally decay into three quarks through the $UDD$ operator. 
In this case the gluinos decay into seven quarks, $\tilde g \to qq L^\prime \to qqqq \tilde \chi_1^0 \to 7 q$, as shown in Fig.~\ref{fig:decay_diag1}.
Note that $\tilde g \to qq \tilde \chi_1^0$ is also kinematically allowed.
However, $Br(\tilde g \to qq L^\prime)/Br(\tilde g \to qq \tilde \chi_1^0)$ is roughly, up to the phase space factor, 
proportional to $|\tilde{\lambda}_{ijk}^{L^\prime} / g|^2$ with $g$ being the electroweak gauge coupling if the neutralino is gaugino-like.
We therefore take $\tilde{\lambda}_{ijk}^{L^\prime} \sim 1$ to suppress the $\tilde g \to qq \tilde \chi_1^0$ mode.

\subsection{Gluino $\to$ nine and more-quarks} 

If one introduces two or more $\bf{5} + \bf{\overline{5}}$ chiral multiplet pairs, even longer gluino decay chains are possible as shown in Fig.~\ref{fig:decay_diag2}.
To enhance the $\tilde g$ and $L'$ decay modes into the heaviest fermionic state possible, we assume a hierarchy in the couplings: 
$\tilde{\lambda}_{1jk}^{L^\prime} > \cdots > \tilde{\lambda}_{njk}^{L^\prime} > g$.
\begin{figure}[t]
  \includegraphics[width=\columnwidth]{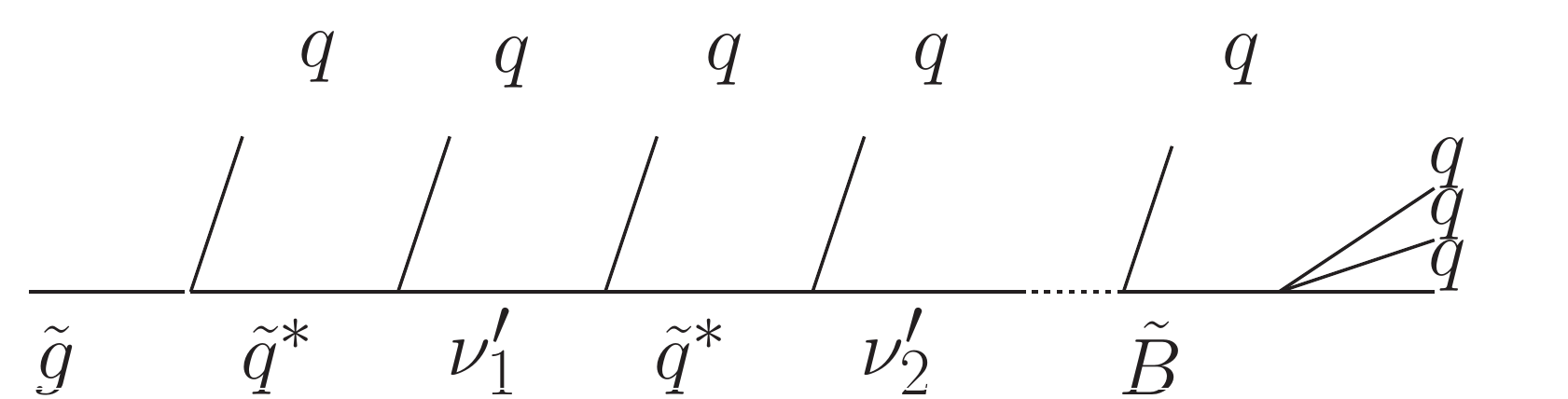}
  \label{fig:decay_diag2}   
  \caption{Possible gluino decay chains which are induced by adding many extra matters with additional RPV term.}
  \label{fig:decay_diag2}     
\end{figure}

As we have seen in this section, the RPV models with extra matters may lead to multi-step gluino decays producing a large number of quarks
and no missing energy in the final state.
In the next section, we reinterpret the existing analyses and study the gluino mass limit in the event topologies considered in this section.

\section{The current LHC constraints}\label{sec:constraints}

As discussed in the previous section, 
in the scenario where several extra matter states involve in the gluino decay chain and the LSP decays hadronically,
the gluino may decay into fully hadronic final states without producing large missing energy.
The event topology with such gluino decay chains is challenging to be searched for at the LHC because of the following reasons:
(1) the standard SUSY searches requiring large $\etmiss$ are not sensitive to this topology.
(2) the final state does not contain isolated leptons, which makes it difficult to distinguish the signal from the backgrounds with fully hadronic final states ($e.g.$ QCD and fully hadronic $t \bar t + {\rm jets}$).
(3) the gluino mass energy is divided into a large number of final state quarks,
making each signal jet soft, which leads to degradation of signal efficiencies because of a high $p_T$ cut threshold for the signal jets. 

The sensitivity of the existing analyses to the models with hadronically decaying gluinos via the baryonic RPV operator has been studied in Ref.~\cite{Evans:2013jna}. 
The authors pointed out that the most stringent constraints were obtained by
the ATLAS 6-7 high $p_T$ jet search \cite{ATLAS:RPVgluino} and the CMS black hole search \cite{Chatrchyan:2013xva}.
The ATLAS 6-7 high $p_T$ jet search looks for excesses in the 6 and 7 exclusive jet multiplicity bins with various $p_T$ cuts:
$> 80$, 100, 120, 140 and 180~GeV.    
The CMS black hole search, on the other hand, employs somewhat smaller $p_T$ cut threshold, 50~GeV.
The analysis uses a kinematic variable, $S_T$, which is defined as the scalar sum of all reconstructed objects, including $\etmiss$,
with $p_T > 50$~GeV.
Ref.~\cite{Evans:2013jna} found that signal events may pollute the control region in the CMS black hole search
and proposed a prescription which assumes the observed data is potentially entirely from signal, with zero background.
This prescription provides conservative limits and we closely follow their analyses.
In particular we use the signal regions ($S_T > 1.9$ and 2.2~TeV and $N_{obj} \ge 8, 9, 10$, where $N_{obj}$ is the number of reconstructed objects, 
excluding $\etmiss$, with $p_T > 50$~GeV) and the corresponding visible cross section upper limits used in Ref.~\cite{Evans:2013jna}. 

In order to estimate the gluino mass bound, we simulate events using {\tt Herwig++} \cite{Bahr:2008pv}.
Generated event samples are then passed to {\tt Delphes} \cite{deFavereau:2013fsa} to simulate detector responses, before estimating signal efficiencies 
for the signal regions.
For the cross section of gluino pair production, we use the values reported by the LHC SUSY cross section working group \cite{SUSYxsecWG},
which includes NLO SUSY QCD corrections and the resummation of soft gluon emission at NLL accuracy.  

\begin{figure}[t]
\centering
  \includegraphics[width=6.5cm]{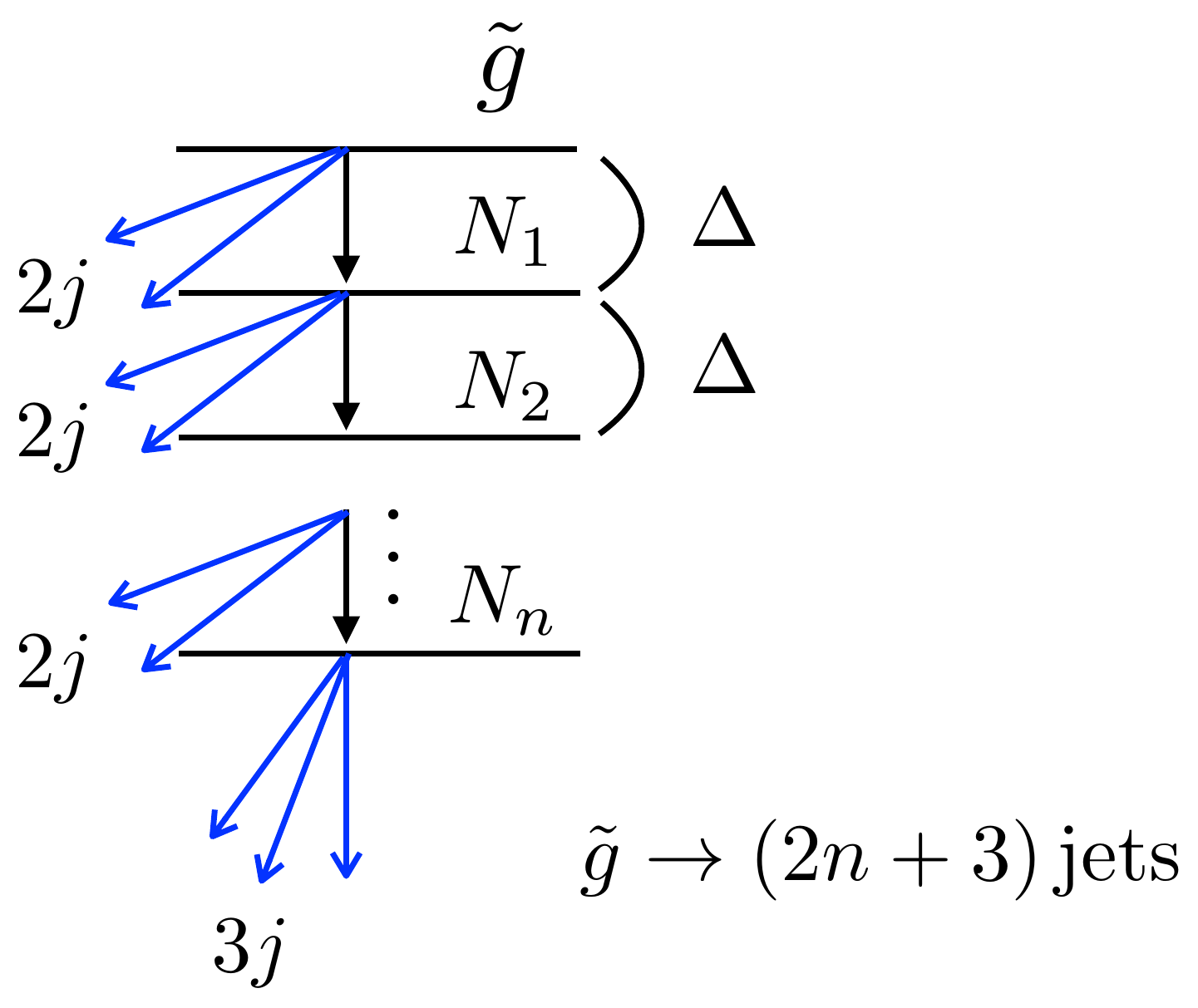}  
  \caption{A generalised gluino decay chain.}
  \label{fig:simptopo} 
\end{figure}

To make complicated gluino decay chains tractable, 
we study the gluino mass limit following simplified model approach. 
In our analysis, we assume squarks are decoupled in the gluino decay and production process, and the gluino decay chain is generalised as a cascade decay
through $n$ intermediate BSM states, $N_1,..., N_n$, with each decay, except for $N_n$, producing two light flavour quarks,
$\tilde g \to qq N_1, N_1 \to qq N_2, ..., N_{n-1} \to qq N_n$, and $N_n$ finally decays into three quarks via the $UDD$ operator, $N_n \to qqq$.
In this setup the gluinos decay into $(2n + 3)$ light flavour quarks. 
In the extra matter scenarios, $N$s are either fermionic extra matter states or the MSSM neutralinos as discussed in the previous section.
For simplicity we assume those BSM states (including gluino) have the same mass gap, $\Delta$, namely 
$m_{\tilde g} - m_{N_1} = m_{N_1} - m_{N_2} =, ..., m_{N_{n-1}} - m_{N_{n}} \equiv \Delta$.
Our generalised gluino decay chain is shown in Fig.~\ref{fig:simptopo}.
 
%
\begin{figure}[t]
  \includegraphics[width=\columnwidth]{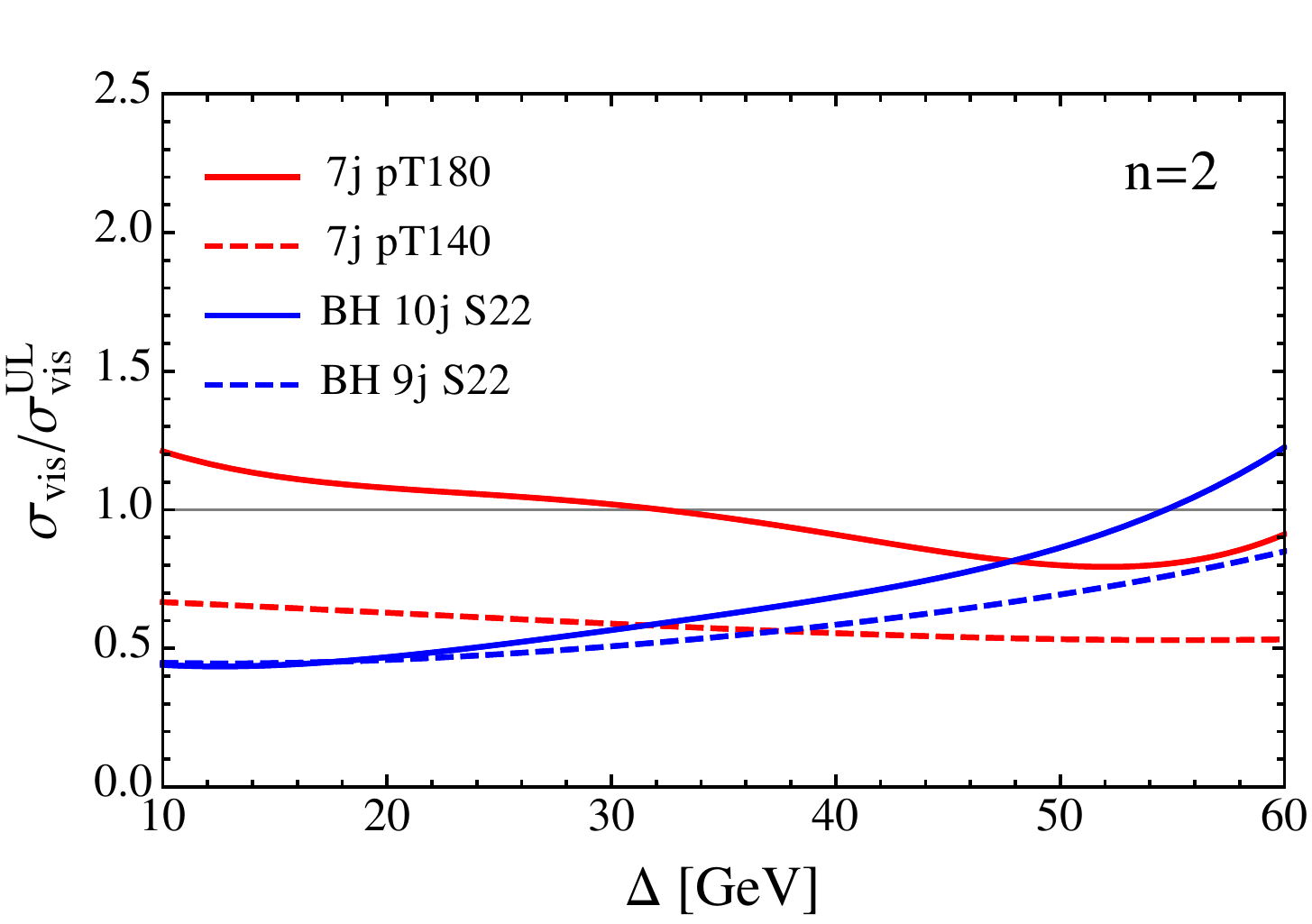}
  \includegraphics[width=\columnwidth]{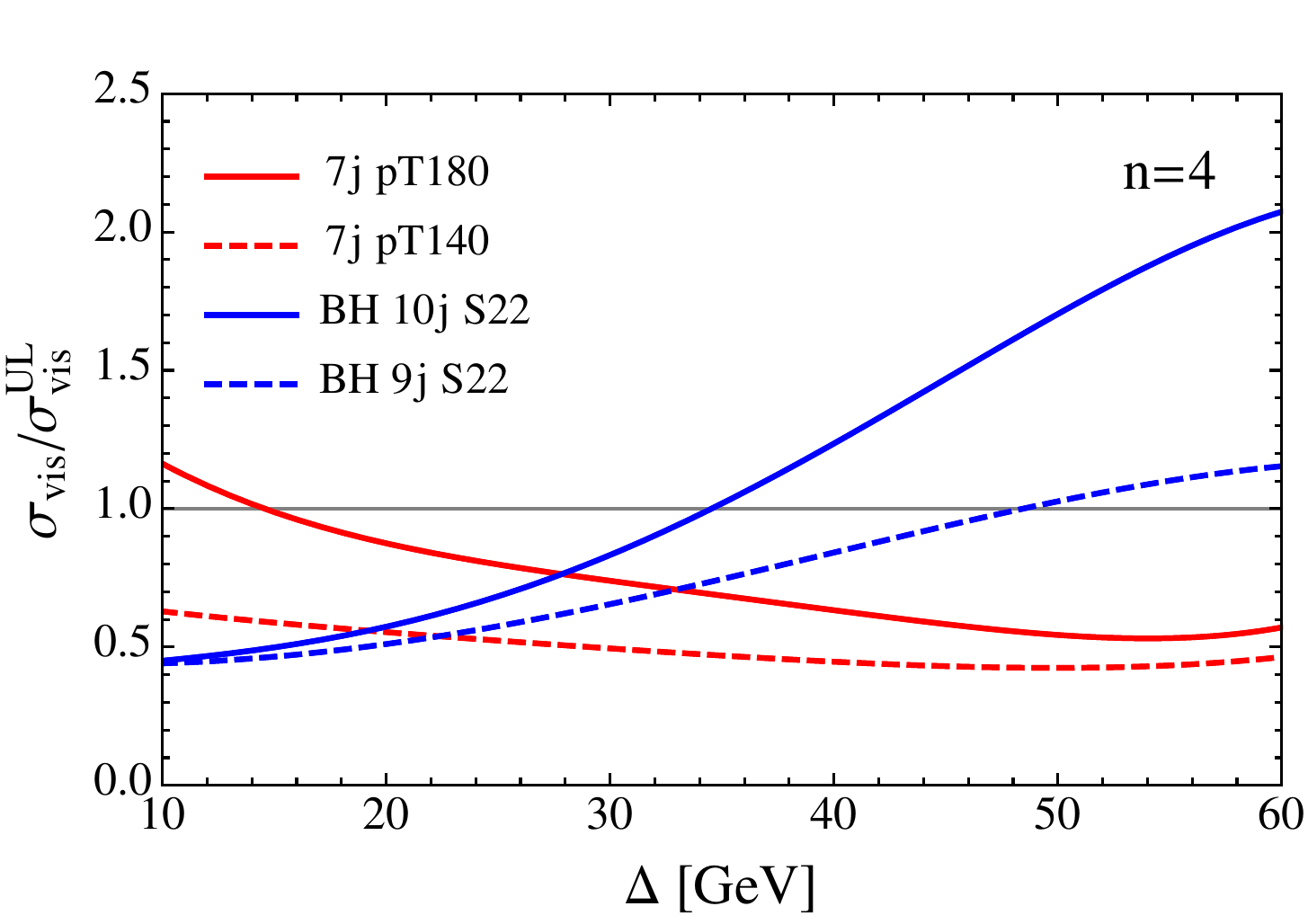}  
  \caption{
  The ratio of the visible cross sections and the 95 \% CL upper limits for the 7 jet with $p_T > 180$ and 140~GeV signal regions by ATLAS (red solid and red dashed) and the 10 and 9 jet with $S_T > 2.2$~TeV signal regions by CMS (blue solid and blue dashed) as functions of the mass gap, $\Delta$.
  The gluino mass is taken at 900~GeV. 
  The upper and lower panels correspond to the models with $n=2$ and 4, respectively.
  }
  \label{fig:delta_sigma_n2} 
\end{figure}
%

Fig.~\ref{fig:delta_sigma_n2} shows the ratio of the visible cross sections and the 95 \% CL upper limits for the 7 jet with $p_T > 180$ and 140~GeV signal regions 
by ATLAS (red solid and red dashed) and the 10 and 9 jet with $S_T > 2.2$~TeV signal regions by CMS (blue solid and blue dashed) as functions of the mass gap, $\Delta$.
The visible cross section is defined as $\sigma_{\rm vis} = \sigma_{\tilde g \tilde g} \cdot \epsilon_i(\Delta)$, where $\sigma_{\tilde g \tilde g}$ is the gluino pair production cross section
and $\epsilon_i(\Delta)$ is the efficiency for signal region $i$, which depends on the mass gap, $\Delta$.  
The gluino mass is taken at 900~GeV and the upper and lower panels correspond to the models with $n=2$ and 4, respectively.
We have checked all the signal regions in 
the ATLAS 6-7 high $p_T$ jet search \cite{ATLAS:RPVgluino} and the CMS black hole search \cite{Chatrchyan:2013xva}
and found that those shown in Fig.~\ref{fig:delta_sigma_n2} provide the strongest constraints.

As can be seen, the constraints obtained from the ATLAS high $p_T$ jet search become weaker as the mass gap $\Delta$ increases up to $50 - 60$~GeV.
This is expected because the $p_T$ scale of the jets coming from $N_n$ decay is characterised by $m_{N_n} = m_{\tilde g} - n \cdot \Delta$.
Contrary, the sensitivity of the CMS black hole search increases as $\Delta$ increases.
This is because the search is sensitive to the events with large jet multiplicity,
and the jets coming from the gluino and $N_m$ ($m < n$) states become hard enough to pass the  $p_T$ cut in this analysis
when $\Delta$ increases.   
Because of the above two effects, a window of the allowed region opens for the 900 GeV gluino at some value of $\Delta$.
For the $n=2$ model, this window appears in the $35 < \Delta/{\rm GeV} < 55$ region.
For the $n=4$ model, these effects are more dramatic compared to the $n=2$ case.
Namely, the ATLAS constrains become weaker and the CMS constraints glows stronger more quickly when increasing $\Delta$ compared to the $n=2$ case.
Consequently, the allowed window for the $n=4$ model appears in the more degenerate mass region: $15 < \Delta/{\rm GeV} < 35.$

\begin{figure}[t]
  \includegraphics[width=\columnwidth]{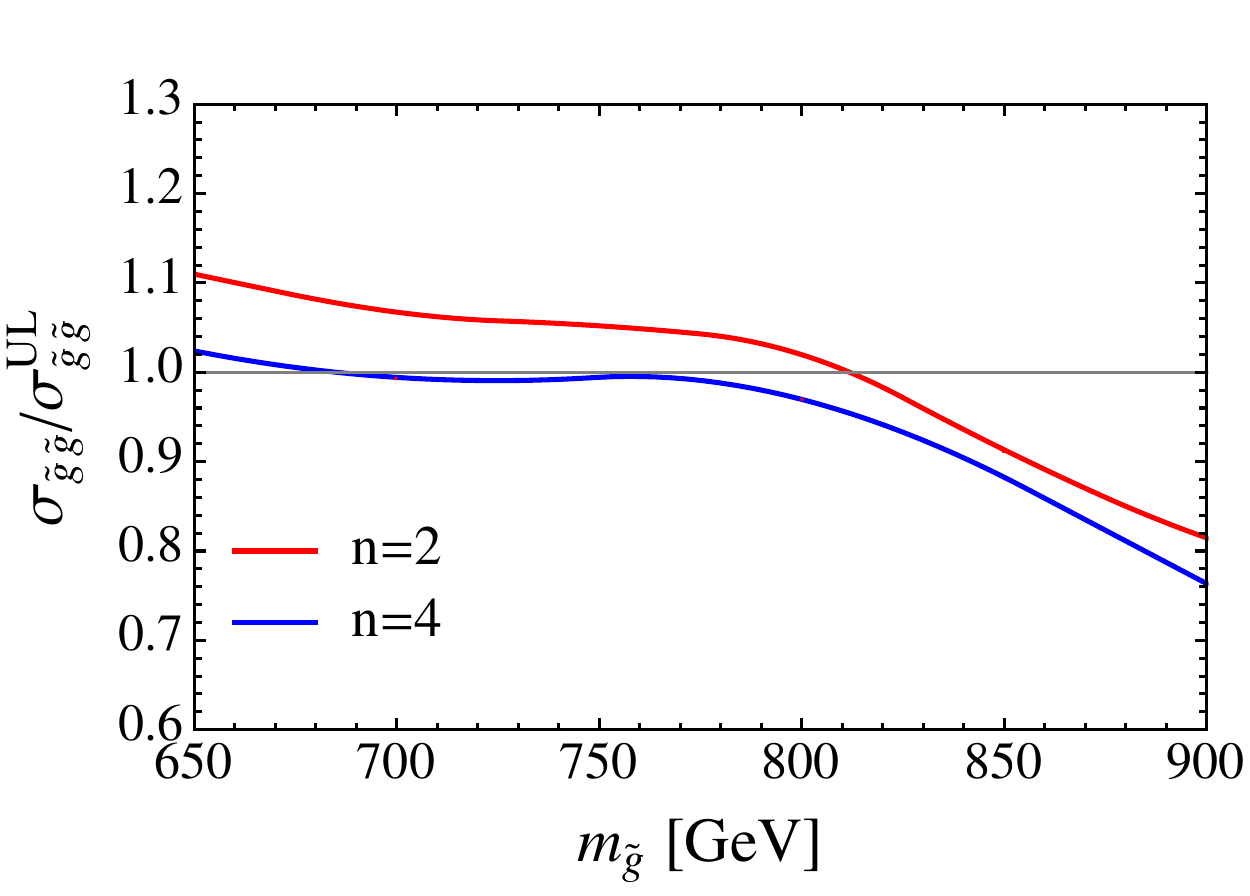}
  \caption{Production cross section constraints for gluino. }
  \label{fig:glimit} 
\end{figure}

In Fig.~\ref{fig:glimit} we show the ratio of the gluino cross section and its 95 \% CL upper limit as a function of the gluino mass.
Here we scan $\Delta$ for each gluino mass and choose the value which gives the weakest sensitivity for that gluino mass. 
We see that the gluino mass limit can be relaxed up to about 810 ($680 - 780$)~GeV for the $n=2$ (4) model. 

%
%

\section{Discussion}\label{sec:discussion}

In the previous section, we found that in the $n=2\, (4)$ model the ATLAS 6-7 jet analysis excludes a 900~GeV gluino only for the $\Delta < 30 \, (15)$~GeV region,
whilst the CMS black hole analysis does so only for the $\Delta > 55 \, (35)$~GeV region.
Consequently a gap in sensitivity arises at the $30 \,(15) < \Delta/{\rm GeV} < 55 \,(35)$ region for the 900~GeV gluino in the $n= 2$\,(4) model.

%
\begin{figure}[t]
  \includegraphics[width=\columnwidth]{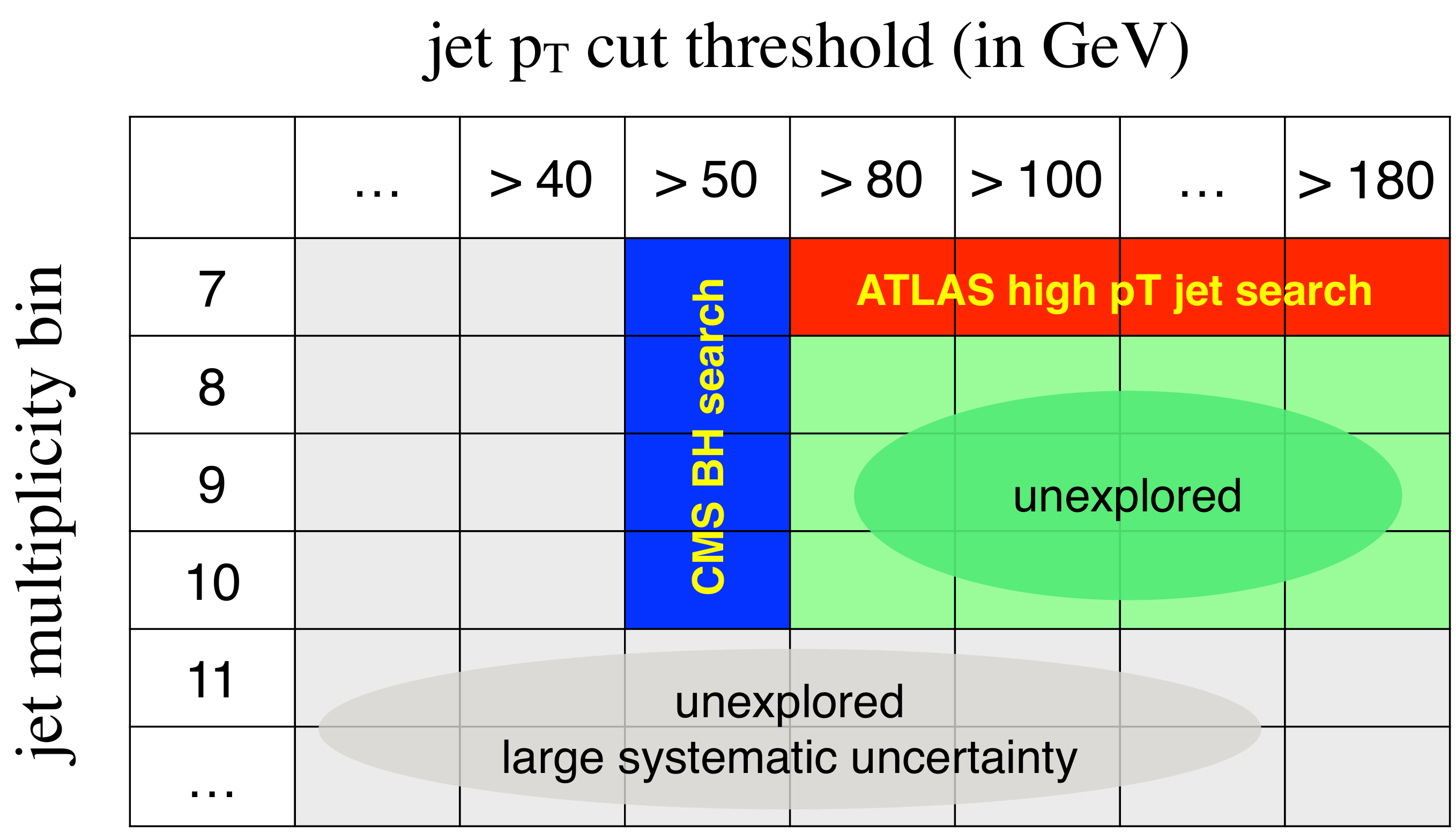}
  \caption{A schematic table for coverage.}
  \label{fig:strategy-table} 
\end{figure}
%

One can expect that this gap can be filled by introducing new signal regions which covers the intermediate region 
between the ATLAS 6-7 jet analysis and the CMS black hole analysis.
Such new signal regions are shown as the green region in Fig.~\ref{fig:strategy-table},
where we summarise the current situation of the analyses by a schematic coverage table.

We would like to comment on the $n \gg 1$ limit.
In this limit the gluino mass energy is divided into a very large number of quarks  
and $p_T$ of each jet would become very small.
Most of the quark jets would fail to pass the 50~GeV jet $p_T$ cut and the analyses with such a $p_T$ cut would not be effective to constrain the model.
To search for $n \gg 1$ scenarios, one would need to extend the search region in the space of the jet $p_T$ cut threshold and the jet multiplicity bin,
as shown in the grey region in Fig.~\ref{fig:strategy-table}.
In this region one has to look at either a very small jet $p_T$ bin or a very large jet multiplicity bin.
It is however challenging to accurately estimate the background contribution to such soft or large jet multiplicity bins. 

We also comment on the CMS gluino resonance search \cite{CMS:TJResonances}. 
This analysis looks for a bump in the three jet invariant mass distribution assuming $\tilde g \to qqq$ topology
using $19.5$ fb$^{-1}$ of $pp$ collision data at $\sqrt{s}=8$ TeV.
Although this is another constraint, we do not expect this analysis changes our result drastically.
In the $n=2$  and 4 models, gluinos decay into 7 and 11 quarks and three quark invariant mass does not reconstruct the gluino mass.
Moreover, because of the large jet multiplicity, the combinatorial background is much larger in the $n=2$ and 4 models,
which would degrade the sensitivity of the analysis significantly.  

The multi-step gluino decay chains can also be realised with the $D^\prime$ intermediate states.
In this case the gluino mass bound would become stronger since $D^\prime$ production events would also contribute to the signal region.
However, in the small $\Delta$ region, the production cross section of $D^\prime$ is smaller than that for the gluinos due to
its smaller colour factor.
The efficiency of the $D^\prime$ production events is also smaller since $D^\prime$ decays fewer quarks compared to the gluinos.
We therefore expect that the gluino mass bound would not change drastically in the models with the $D^\prime$ intermediate states.
If $L^\prime$ are also included in this system with $m_{\tilde g} > m_{D^\prime} > m_{L^\prime}$, 
the gluino decay chain would become longer and it may be helpful to relax the gluino mass bound.

We also comment on the models with a ${\bf 10} + \bf{\overline{10}}$ pair, where ${\bf 10} = (Q^\prime, U^\prime, E^\prime$) and ${\bf \overline{10}} = (\overline{Q}^\prime, \overline{U}^\prime, \overline{E}^\prime$). 
The multi-step gluino decay chains via the $U^\prime$ intermediate states are possible in this model using  the $U^\prime DD$ operator.
Finally, it is worth pointing out that in our scenario the mass bound on the extra matters are also relaxed since 
the extra matter states decay fully hadronically through long cascade decay chains.

\section{Conclusions}\label{sec:conclusion}

In this paper we pointed out that the RPV models with extra matters may lead to multi-step gluino decays into fully hadronic final states.
We reinterpreted the existing analyses and studied the gluino mass bound in our generalised model.
In the region where the mass gap $\Delta$ is small, the sensitivity of ATLAS 6-7 jets analysis decreases, whilst that of CMS black hole search increases
as increasing $\Delta$.
Consequently, the current LHC sensitivity is minimised at some value of $\Delta$. 
In the simulation we demonstrated that the gluino mass bound in such scenarios can be as small as $700$ or $800$~GeV 
depending on the number of intermediate states in the gluino decay chain. 

In order to increase the sensitivity to the gluinos that undergo multi-step cascade decays into fully hadronic final states,
it is important to extend the search strategy in the space of the jet $p_T$ cut threshold and the jet multiplicity bin,
which requires a better understanding of the backgrounds contributing the soft and large jet multiplicity bins.


\paragraph{Acknowledgments} 

M.A. acknowledges support from the German Research Foundation (DFG) through
grant BR 3954/1-1 and DFG TRR33 "The Dark Universe". 
The work of K.S. was supported in part by
the London Centre for Terauniverse Studies (LCTS), using funding from
the European Research Council 
via the Advanced Investigator Grant 267352.
The work of T.T.Y. was supported by 
Scientific Research (B) No. 26287039 [TTY]
and
World Premier International Center Initiative
(WPI Program), MEXT, Japan.


\end{document}